\documentclass[onecolumn]{elsarticle}

\usepackage{amsmath}
\usepackage{amssymb}
\usepackage{graphicx}
\usepackage{epsfig}
\usepackage{ulem}
\usepackage{cancel}
\usepackage{color}

\def\comment#1{}

\journal{Physics Letters B}

\begin{document}

\begin{frontmatter}

\title{The Klein first integrals in an equilibrium system with electromagnetic, weak, strong and gravitational interactions}

\author[roma,pescara]{Jorge A. Rueda}
\ead{jorge.rueda@icra.it}

\author[roma,pescara,nizza]{R. Ruffini}
\ead{ruffini@icra.it}

\author[roma,pescara]{S.-S. Xue}
\ead{xue@icra.it}

\address[roma]{Dipartimento di Fisica and ICRA, Sapienza Universita' di Roma\\
P.le Aldo Moro 5, I-00185 Rome, Italy}
\address[pescara]{ICRANet, P.zza della Repubblica 10, I-65122 Pescara, Italy}
\address[nizza]{ICRANet, University of Nice-Sophia Antipolis, 28 Av. de Valrose, 06103 Nice Cedex 2, France}

\begin{abstract}
The isothermal Tolman condition and the constancy of the Klein potentials originally expressed for the sole gravitational interaction in a single fluid are here generalized to the case of a three quantum fermion fluid duly taking into account the strong, electromagnetic, weak and gravitational interactions. The set of constitutive equations including the Einstein-Maxwell-Thomas-Fermi equations as well as the ones corresponding to the strong interaction description are here presented in the most general relativistic isothermal case. This treatment represents an essential step to correctly formulate a self-consistent relativistic field theoretical approach of neutron stars.
\end{abstract}

\begin{keyword} 
Neutron Star Matter; General Relativity; Thermodynamics of General Relativistic Systems.
\end{keyword}

\end{frontmatter}

\section{Introduction}\label{sec:intro}	
%
The unsolved problems of supernovae theories as well as the necessity of processes leading to electrodynamical phenomena during the gravitational collapse to a black hole \cite{physrep} lead to the necessity of critically reexamining the current treatment of neutron stars. In a series of articles (see \cite{PRC2011,PLB2011}), we have recently developed the first steps towards a new consistent treatment for the description of neutron stars, well beyond the traditional Tolman-Oppenheimer-Volkoff equations. 

First we have generalized the treatment of compressed atoms of Feynman, Metropolis and Teller to the relativistic regimes (see \cite{PRC2011} for details). There, it has been enforced self-consistently in a relativistic Thomas-Fermi equation, the condition of $\beta$-equilibrium extending the works of V.~S.~Popov \cite{popov71}, Ya.~B.~Zeldovich and V.~S.~Popov \cite{zeldovich72}, A.~B.~Migdal et al.~\cite{migdal76,migdal77}, J.~Ferreirinho et al.~\cite{ferreirinho80} and R.~Ruffini and L.~Stella \cite{ruffini81} for heavy nuclei. Then, through the using of scaling laws, following \cite{dresden,popovGX,popovminsk}, this treatment was extrapolated to compressed nuclear matter cores at nuclear and supranuclear densities. Such cores have stellar dimensions and mass numbers $A\simeq (m_{\rm Planck}/m_n)^3 \sim 10^{57}$ or $M_{\rm core}\sim M_{\odot}$. In addition, they fulfill global but not local charge neutrality having electric fields on the core surface, increasing for decreasing values of the electron Fermi energy $E^F_e$ reaching values much larger than the critical value $E_c = m_e^2c^3/(e\hbar)$, for $E_e^F=0$. The assumption of constant distribution of protons at nuclear densities simulates, in such a treatment, the confinement due to the strong interactions in the case of nuclei and heavy nuclei and due to both the gravitational field and strong interactions in the case of nuclear matter cores of stellar sizes at nuclear and supranuclear densities.

In a subsequent work \cite{PLB2011}, we have generalized the above approach explicitly including the effects of the gravitational field by considering the most simplified nontrivial but rigorous treatment of a general relativistic system of neutrons, protons and electrons in $\beta$-equilibrium. It has been there proved that the traditional treatment for the description of neutron stars adopting the condition of local charge neutrality is not consistent with the Einstein-Maxwell equations and with microphysical conditions of equilibrium within quantum statistics. The role of the constancy of the general relativistic Fermi energy of each particle species pioneered by O.~Klein \cite{klein49} has been there emphasized and, the full system of equilibrium equations consisting of the Einstein-Maxwell and general relativistic Thomas-Fermi equations has been formulated. The corresponding solution of such a system of equations has been there given in the simplest possible example of a configuration of neutrons, protons and electrons in $\beta$-equilibrium with electromagnetic, weak and gravitational interactions. New electrodynamic and general relativistic properties of the equilibrium configurations have been there illustrated.

The aim of this article is to make an essential new step: we further proceed to the description of a system of neutrons, protons and electrons fulfilling strong, electromagnetic, weak and gravitational interactions. The essential role of the Klein first integrals is evidenced and their theoretical formulation is presented in the Einstein-Maxwell background. For the sake of generality the treatment is performed in the most general case in which finite temperature effects are also taking into account. We adopt throughout the article natural units $\hbar=c=1$.

\section{The Constitutive General Relativistic Equations}\label{sec:GFor}	

The densities in the core of a neutron star exceed the nuclear density $\rho_{\rm nuc} \sim 2.7 \times 10^{14}$ g/cm$^3$ and may reach densities of order $\sim 10^{17}$ g/cm$^3$ at the verge of the gravitational collapse of the neutron star to a black hole. There is therefore the need of a consistent relativistic theory for the description of the interactions between the matter constituents. In particular, approaches for the nuclear interaction between nucleons based on phenomenological potentials and non-relativistic many-body theory become inapplicable (see \cite{bowers73a,bowers73b}). 

A self-consistent relativistic and well-tested model for the nuclear interactions is the Walecka model (see \cite{duerr56,walecka74} for details). This model share common features with the model adopted by Bowers et al. in \cite{bowers73a,bowers73b}; in both of them the nucleons interact through a Yukawa coupling and the flat spacetime has been considered to construct the equation of state of nuclear matter. The technique of constructing the equation of state assuming flat spacetime has been generally used since, as pointed out in \cite{bowers73a,bowers73b}, as long as $\rho < 10^{49}$ g/cm$^3$ the gravitational contributions to interactions between particles are negligible. However, when we turn to neutron star configurations at nuclear and supranuclear densities, it has been shown in \cite{PLB2011} how the solution of the Einstein-Maxwell system of equations is mandatory.

In the often called extended version of the Walecka model, the strong interaction between nucleons is described by the exchange of three virtual mesons: $\sigma$ is an isoscalar meson field providing the attractive long-range part of the nuclear force; $\omega$ is a massive vector field that models the repulsive short range and; $\rho$ is the massive isovector field that takes account surface as well as isospin effects of nuclei (see also \cite{boguta77,ring96}).

The total Lagrangian density of the system is given by 
\begin{equation}\label{Eq:TLD}
\mathcal{L}=\mathcal{L}_{g}+\mathcal{L}_{f}+\mathcal{L}_{\sigma}+\mathcal{L}_{\omega}+\mathcal{L}_{\rho}+\mathcal{L}_{\gamma}+\mathcal{L}_{\rm int},
\end{equation}
where the Lagrangian densities for the free-fields are
\begin{eqnarray}
\mathcal{L}_g &=& -\frac{R}{16 \pi G},  \label{eq:Lg}\\
\mathcal{L}_{\gamma} &=& -\frac{1}{4} F_{\mu\nu}F^{\mu\nu},  \label{eq:Lgamma}\\
\mathcal{L}_{\sigma} &=& \frac{1}{2} \nabla_{\mu}\sigma \nabla^{\mu}\sigma-U(\sigma),  \label{eq:Ls}\\
\mathcal{L}_{\omega} &=& -\frac{1}{4} \Omega_{\mu\nu}\Omega^{\mu\nu}+\frac{1}{2}m_{\omega}^{2} \omega_{\mu} \omega^{\mu},  \label{eq:Lomega} \\
\mathcal{L}_{\rho} &=& -\frac{1}{4} \mathcal{R}_{\mu\nu}\mathcal{R}^{\mu\nu}+\frac{1}{2}m_{\rho}^{2} \rho_{\mu} \rho^{\mu},   \label{eq:Lrho}
\end{eqnarray}
where $\Omega_{\mu\nu}\equiv\partial_{\mu}\omega_{\nu}-\partial_{\nu}\omega_{\mu}$, $\mathcal{R}_{\mu\nu}\equiv\partial_{\mu}\rho_{\nu}-\partial_{\nu}\rho_{\mu}$,
$F_{\mu\nu}\equiv\partial_{\mu}A_{\nu}-\partial_{\nu}A_{\mu}$ are the field strength tensors for the
$\omega^{\mu}$, $\rho$ and $A^{\mu}$ fields respectively, $\nabla_\mu$ stands for covariant derivative and $R$ is the Ricci scalar. We adopt the Lorenz gauge for the fields $A_\mu$, $\omega_\mu$, and $\rho_\mu$. The self-interaction scalar field potential $U(\sigma)$ is a quartic-order polynom for a renormalizable theory (see e.g.~\cite{lee74}). The specific functional form of $U(\sigma)$ is not relevant for the scope of this article, thus we will not adopt any particular form of it hereafter.

The Lagrangian density for the three fermion species is
\begin{equation}
\mathcal{L}_f = \sum_{i=e, N}\bar{\psi}_{i}\left(i \gamma^\mu D_\mu-m_i \right)\psi_i, \label{eq:Lf}
\end{equation}
where $\psi_N$ is the nucleon isospin doublet, $\psi_e$ is the electronic singlet, $m_i$ states for the mass of each particle-specie and $D_\mu = \partial_\mu + \Gamma_\mu$, being $\Gamma_\mu$ the Dirac spin connections that satisfy the commutation relation
\begin{equation}
[\gamma_\mu, \Gamma_\nu] = \partial_\nu \gamma_\mu -\Gamma^\alpha_{\mu \nu} \gamma_\alpha, 
\end{equation}
where $\Gamma^{\alpha}_{\mu \nu}$ denotes the Christoffel symbols.

The interacting part of the Lagrangian density is, in the minimal coupling assumption, given by
\begin{equation}\label{eq:Lint}
\mathcal{L}_{\rm int} = -g_{\sigma} \sigma \bar{\psi}_N \psi_N - g_{\omega} \omega_{\mu} J_{\omega}^{\mu}-g_{\rho}\rho_{\mu}J_{\rho}^{\mu}+e A_{\mu} J_{\gamma,e}^{\mu}-e A_{\mu} J_{\gamma,N}^{\mu},
\end{equation}
where the conserved currents are
\begin{eqnarray}
J^{\mu}_{\omega} &=& \bar{\psi}_N \gamma^{\mu}\psi_N, \label{eq:J1}\\
J^{\mu}_{\rho} &=& \bar{\psi}_N \tau_3\gamma^{\mu}\psi_N, \label{eq:J2}\\
J^{\mu}_{\gamma, e} &=& \bar{\psi}_e \gamma^{\mu}\psi_e, \label{eq:J3}\\
J^{\mu}_{\gamma, N} &=& \bar{\psi}_N \left(\frac{1+\tau_3}{2}\right)\gamma^{\mu}\psi_N. \label{eq:J4}
\end{eqnarray}

The coupling constants of the $\sigma$, $\omega$ and $\rho$-fields are $g_{\sigma}$, $g_{\omega}$ and $g_{\rho}$, and $e$ is the fundamental electric charge. The  Dirac matrices $\gamma^{\mu}$ and the isospin Pauli matrices satisfy the Dirac algebra in curved spacetime (see e.g.~\cite{lee87})
\begin{eqnarray}
\label{Eq:1D}
\left\{\gamma^{\mu}, \gamma^{\nu}\right\} &=& 2g^{\mu\nu}, \\
\left\{\gamma_{\mu}, \gamma_{\nu}\right\} &=& 2g_{\mu\nu}, \\
\left\{\gamma^{\mu}, \gamma_{\nu}\right\} &=& 2\delta^{\mu}_{\nu}, \\
\label{Eq:4D}
\left[\tau_i, \tau_j\right] &=& 2 \imath \epsilon_{i j k} \tau^k.
\end{eqnarray}

The Einstein-Maxwell-Dirac system of equations is then given by
\begin{eqnarray}
G_{\mu\nu} + 8 \pi G T_{\mu\nu} &=& 0, \label{eq:EM1}\\
\nabla_{\mu}F^{\mu\nu}- e J_{ch}^{\nu} &=& 0, \label{eq:EM2}\\
\nabla_{\mu}\Omega^{\mu\nu}+m_{\omega}^2 \omega^{\nu}-g_{\omega}J_{\omega}^{\nu} &=& 0, \label{eq:EM6}\\
\nabla_{\mu}\mathcal{R}^{\mu\nu}+m_{\rho}^2 \rho^{\nu}- g_{\rho}J_{\rho}^{\nu} &=& 0, \label{eq:EM7}\\
\nabla_{\mu}\nabla^{\mu} \sigma+\partial_{\sigma}U(\sigma)+g_s n_s &=& 0, \label{eq:EM5}\\
\left[\gamma_{\mu}\left(i D^\mu - V^\mu_N\right) - \tilde{m}_N\right]\psi_N &=& 0, \label{eq:EM3}\\
\left[\gamma_{\mu}\left(i D^\mu + e A^{\mu}\right) - m_e\right]\psi_e &=& 0, \label{eq:EM4}
\end{eqnarray}
where the scalar density $n_s = \bar{\psi}_N \psi_N$, the nucleon effective mass $\tilde{m}_N\equiv m_N + g_{\sigma}\sigma$, and
\begin{eqnarray}
V^\mu_N\equiv g_{\omega}\omega^{\mu}+ g_{\rho}\tau \rho^{\mu}+e \left(\frac{1+\tau_3}{2}\right) A^{\mu},
\label{eq:VN}
\end{eqnarray}
is the effective four potential of nucleons.  The energy-momentum tensor of free-fields and free-fermions $T^{\mu\nu}$ of the system (\ref{eq:Lgamma})--(\ref{eq:Lrho}) is
\begin{equation}\label{eq:Tab}
T^{\mu \nu} =  T_{f}^{\mu\nu} + T_{\gamma}^{\mu\nu} + T_{\sigma}^{\mu\nu} + T_{\omega}^{\mu\nu} + T_{\rho}^{\mu\nu},
\end{equation}
where
\begin{eqnarray}
T_{\gamma}^{\mu\nu} &=& -F_{\alpha}^{\mu}F^{\alpha \nu}-\frac{1}{4}g^{\mu\nu}F_{\alpha\beta}F^{\alpha\beta},
\\
T_{\sigma}^{\mu\nu} &=& \nabla^{\mu}\nabla^{\nu}\sigma-g^{\mu\nu}\left[\frac{1}{2}\nabla_{\sigma}\sigma\nabla^{\sigma}\sigma-U(\sigma)\right],\\
T_{\omega}^{\mu\nu} &=& -\Omega_{\alpha}^{\mu}\Omega^{\alpha\nu}-\frac{1}{4}g^{\mu\nu}\Omega_{\alpha\beta}\Omega^{\alpha\beta}+m_{\omega}^2\left(\omega^{\mu}\omega^{\nu}-\frac{1}{2}g^{\mu\nu}\omega_{\alpha}\omega^{\alpha}\right),\\
T_{\rho}^{\mu\nu} &=& -\mathcal{R}_{\alpha}^{\mu}\mathcal{R}^{\alpha\nu}-\frac{1}{4}g^{\mu\nu}\mathcal{R}_{\alpha\beta}\mathcal{R}^{\alpha\beta}+m_{\rho}^2\left(\mathcal{R}^{\mu}\mathcal{R}^{\nu}-\frac{1}{2}g^{\mu\nu}\mathcal{R}_{\alpha}\omega^{\alpha}\right),
\end{eqnarray}
are the contribution due to free-fields and $T_{f}^{\mu\nu}$ is the contribution of free-fermions which we discuss below.

\section{The Thermodynamic Laws and the Field Equations in the Spherically Symmetric Case}\label{sec:thermodynamics}	

We first introduce the non-rotating spherically symmetric spacetime metric
\begin{equation}\label{Eq:metriSN}
ds^2= {\rm e}^{\nu(r)}dt^2-{\rm e}^{\lambda(r)}dr^2-r^2d\theta^2 - r^2 \sin^2\theta d\varphi^2,
\end{equation}
where the $\nu(r)$ and $\lambda(r)$ are only functions of the radial coordinate $r$.

For very large number of fermions, we can adopt the mean-field approximation in which fermion-field operators are replaced by their expectation values (see e.g.~\cite{walecka74} for details). 

We write the nucleon doublet and the electronic spinor as $\psi_i=\psi_i(k){\rm e}^{-i k_{\mu}x^{\mu}}$ in the phase-space. Suppose that neutrons, protons and electrons, and the corresponding antiparticles, are in thermodynamic equilibrium with a finite temperature $T$. The occupation fermion-number operators of the ``$k$''-state, ${\mathcal N}_i(k)=\psi^\dagger_i(k)\psi_i(k)$ with $i=e,p,n$, are replaced by their Fermi-distributions
\begin{equation}\label{eq:dis_k}
f^\pm_i(k) = \langle \psi^\pm_i(k)^\dagger\psi^\pm_i(k)\rangle = \left[\exp\left(\frac{\epsilon_i(k)\mp \mu_i}{k_B T}\right)+1\right]^{-1},
\end{equation}
where $k_B$ is the Boltzmann constant, $\mu_i$ and $\epsilon_i(k)=\sqrt{k^2+\tilde{m}^2_i}$ denote the single-particle chemical potential and energy-spectrum (we recall that for electrons $\tilde{m}_e = m_e$). The sign `$+$' correspond to particles and `$-$' to antiparticles. We do not consider ``real'' bosons to be present in the system; the only distribution functions involved in the computation are due to fermions and antifermions and therefore phenomena as Bose-Einstein condensation does not occur within this theory (see e.g.~\cite{bowers73a} for details).

It is worth to recall that all the thermodynamic quantities, e.g.~$k$, $\epsilon$, $T$ ..., are written here in the local frame which is related to the coordinate frame by the Lorentz ``boost''
\begin{equation}
\Lambda^{(a)}_\alpha = (u_\alpha,\chi_\alpha,\Theta_\alpha,\Phi_\alpha),
\end{equation}
where $u_\alpha = e^{\nu/2} \delta^0_\alpha$, $\chi_\alpha = e^{\lambda/2} \delta^1_\alpha$, $\Theta_\alpha=r \delta^2_\alpha$, and $\Phi_\alpha=r \sin\theta \delta^3_\alpha$, being $\delta^\alpha _\beta$ the usual Kronecker delta symbol.

The number-density $n_i$ of the $i$-specie, taking into account the antiparticle contribution is, within the mean-field approximation, given by
\begin{equation}
n_i=\frac{2}{(2 \pi)^3}\int d^3k [f^+_i(k)-f^-_i(k)].
\label{eq:dis}
\end{equation}
The contribution of free-fermions and antifermions to the energy-momentum tensor can be then written in the perfect fluid form (see e.g.~\cite{ruffini69})
\begin{equation}
T_{f}^{\mu\nu} = (\mathcal{E}+\mathcal{P})u^{\mu}u^{\nu}-\mathcal{P}g^{\mu\nu},
\end{equation}
where $u^\mu$ is the four-velocity of the fluid which satisfies $u^\mu u_\mu=1$, and the energy-density $\mathcal{E}$ and the pressure $\mathcal{P}$ are given by
\begin{equation}\label{eq:EOS1}
\mathcal{E}=\sum_{i=n,p,e}\mathcal{E}_i, \qquad \mathcal{P}=\sum_{i=n,p,e}\mathcal{P}_i,
\end{equation}
being $\mathcal{E}_i$ and $\mathcal{P}_i$ the single fermion-antifermion fluid contributions
\begin{eqnarray}
\mathcal{E}_i &=&\frac{2}{(2 \pi)^3} \int d^3k\epsilon_i(k) [f^+_i(k)+f^-_i(k)], \label{eq:EOS2a}\\
\mathcal{P}_i&=&\frac{1}{3}\frac{2}{(2 \pi)^3} \int d^3k \frac{k^2} {\epsilon_i(k)} [f^+_i(k)+f^-_i(k)]. \label{eq:EOS2b}
\end{eqnarray}
The equation of state (\ref{eq:EOS1})--(\ref{eq:EOS2b}) satisfies the thermodynamic law
\begin{equation}\label{eq:thermolaw}
\mathcal{E}+\mathcal{P}-T {\cal S} =\sum_{i=n,p,e}n_i \mu_i,
\end{equation}
where ${\cal S} = S/V$ is the entropy per unit volume (entropy density) and $\mu_i= \partial\mathcal{E}/\partial n_i$ is the free-chemical potential of the $i$-specie. At zero-temperature $T=0$, $\mu_i =\sqrt{(K_i^F)^2+\tilde{m}^2_i}$ and $n_i=(K_i^F)^3/(3 \pi^2)$, where $K_i^F$ denotes the Fermi momentum of the $i$-specie.

The scalar density $n_s$, within the mean-field approximation, is given by the following expectation value
\begin{equation}\label{eq:ns}
n_s = \langle\bar{\psi}_N \psi_N\rangle = \frac{2}{(2 \pi)^3}\sum_{i=n,p}\int d^3k\frac{\tilde{m}_N}{\epsilon_{i}(k)}[f^+_i(k)+f^-_i(k)].
\end{equation}

In the static case, only the temporal components of the covariant currents survive, i.e. $\langle \bar{\psi}(x)\gamma^{i}\psi(x)\rangle=0$. Thus, by taking the expectation values of Eqs.~(\ref{eq:J1})--(\ref{eq:J4}), we obtain the non-vanishing components of the currents 
\begin{eqnarray}
J^{ch}_0 &=& n_{ch} u_0=(n_p-n_e)u_0, \label{eq:J12}\\
J^{\omega}_0 &=& n_b u_0=(n_n+n_p)u_0, \label{eq:J22}\\
J^{\rho}_0 &=& n_3 u_0=(n_p-n_n)u_0, \label{eq:J32}
\end{eqnarray}
where $n_b$, $n_p$, $n_n$  and $n_e$ are the baryon, proton, neutron and electron number densities which are functions only of the spatial coordinates, and $u_0 = \sqrt{g_{00}} = e^{\nu/2}$.

Making a variation of Eq.~(\ref{eq:thermolaw}) and using Eqs.~(\ref{eq:EOS1})--(\ref{eq:EOS2b}) and (\ref{eq:ns}), we obtain the generalized Gibbs-Duhem relation
\begin{equation}\label{Eq:GGDR}
d\mathcal{P}=\sum_{i=n,p,e} n_i d\mu_i -g_{\sigma}n_s d\sigma + {\cal S} dT,
\end{equation}
which can be rewritten as
\begin{equation}\label{Eq:GGDR2}
d\mathcal{P}=\sum_{i=n,p,e} n_i d\mu_i -g_{\sigma}n_s d\sigma + \left({\cal E}+{\cal P}-\sum_{i=n,p,e} n_i \mu_i \right) \frac{dT}{T},
\end{equation}
where we have used Eq.~(\ref{eq:thermolaw}) to eliminate ${\cal S}$, and we have used the relation between the scalar density and the fluid energy-density
\begin{equation}
n_s = \frac{\partial \mathcal{E}}{\partial\tilde{m}_N},
\end{equation}
which follows from Eqs.~(\ref{eq:EOS1})--(\ref{eq:EOS2b}) and (\ref{eq:ns}).

Therefore, the Einstein-Maxwell equations (\ref{eq:EM1})--(\ref{eq:EM5}), within the mean-field approximation, become
\begin{eqnarray}
&&{\rm e}^{-\lambda(r)}\left(\frac{1}{r^2}-\frac{1}{r}\frac{d\lambda}{dr}\right)-\frac{1}{r^2}= -8 \pi G T_0^0, \label{eq:EM12}\\
&&{\rm e}^{-\lambda(r)}\left(\frac{1}{r^2}+\frac{1}{r}\frac{d\nu}{dr}\right)-\frac{1}{r^2}= -8 \pi G T_1^1, \label{eq:EM22}\\
&&{\rm e}^{-\lambda(r) }\left[\frac{1}{2}\left(\frac{d\nu}{dr}-\frac{d\lambda}{dr}\right)\left(\frac{1}{r}+ \frac{1}{2}\frac{d\nu}{dr}\right)+\frac{1}{2}  \frac{d^2\nu}{dr^2}\right]= -8 \pi G T_3^3, \label{eq:EMpiu}\\
&&\frac{d^2V}{dr^2}+\frac{dV}{dr} \left[\frac{2}{r}-\frac{1}{2}\left(\frac{d\nu}{dr}+\frac{d\lambda}{dr}\right)\right]=-{\rm e}^{\lambda}e J_{ch}^0, \label{eq:EM32}\\
&&\frac{d^2\sigma}{dr^2}+\frac{d\sigma}{dr}\left[\frac{2}{r}-\frac{1}{2}\left(\frac{d\nu}{dr}+\frac{d\lambda}{dr}\right)\right]={\rm e}^{\lambda}\left[\partial_{\sigma}U(\sigma)+g_s n_s\right], \label{eq:EM42}\\
&&\frac{d^2\omega}{dr^2}+\frac{d\omega}{dr}\left[\frac{2}{r}-\frac{1}{2}\left(\frac{d\nu}{dr}+\frac{d\lambda}{dr}\right)\right]=-{\rm e}^{\lambda}\left[g_{\omega}J_{\omega}^0-m_{\omega}^2\omega\right], \label{eq:EM52}\\
&&\frac{d^2\rho}{dr^2}+\frac{d\rho}{dr}\left[\frac{2}{r}-\frac{1}{2}\left(\frac{d\nu}{dr}+\frac{d\lambda}{dr}\right)\right]=-{\rm e}^{\lambda}\left[g_{\rho} J_{\rho}^0-m_{\rho}^2\rho\right], \label{eq:EM62}
\end{eqnarray}
where we have introduced the notation $\omega_0=\omega$, $\rho_0=\rho$, and $A_0=V$. The metric function $\lambda$ is related to the mass $M(r)$ and the electric field $E(r) = -{\rm e}^{-(\nu+\lambda)/2} V'$ through 
\begin{equation}\label{eq:lambda}
{\rm e}^{-\lambda(r)} = 1 - \frac{2 G M(r)}{r} + G r^2 E^2(r) = 1 - \frac{2 G M(r)}{r} + \frac{G Q^2(r)}{r^2}\, ,
\end{equation}
where we have introduced also the conserved charge $Q(r)=r^2 E(r)$.

An important equation, although not independent of the Einstein-Maxwell equations (\ref{eq:EM12})--(\ref{eq:EM62}), is given the energy-momentum conservation law 
\begin{equation}
\nabla_\mu T^{\mu \nu}= - g_{\omega} J^{\omega}_{\mu} \Omega^{\mu \nu}-g_{\rho} J^{\rho}_{\mu} {\cal R}^{\mu \nu} + e J^{ch}_{\mu} F^{\mu \nu},
\end{equation}
from which we have
\begin{equation}\label{eq:TOV}
\frac{d\mathcal{P}}{dr} = - \frac{(\mathcal{E}+\mathcal{P})}{2}\frac{d\nu}{dr} -g_{\sigma}n_s \frac{d\sigma}{dr} - g_{\omega}J_{\omega}^0\frac{d\omega}{dr} -  g_{\rho} J_{\rho}^0\frac{d\rho}{dr} - e J^0_{ch}\frac{dV}{dr},
\end{equation}
where we have used the energy-momentum tensor $T^{\mu \nu}$ given by Eq.~(\ref{eq:Tab}).
%

\section{Constancy of the Klein potentials and $\beta$-equilibrium}\label{sec:COTGFE}

Introducing the nucleon doublet and the electronic spinor in the wave-form $\psi_i=\psi_i(k){\rm e}^{-i k_{\mu}x^{\mu}}$ in phase-space, the Dirac equations (\ref{eq:EM4}) become
\begin{equation}
(\gamma_{\mu}\mathcal{K}^{\mu}_i-\tilde{m}_i)\psi_i(k)=0,
\label{eq:dirac}
\end{equation}
where
\begin{equation}
\mathcal{K}^{\mu}_i\equiv k^{\mu}-V_i^{\mu},\quad  V_e^{\mu}=-eA^{\mu}.
\end{equation}
In the mean-field approximation, making the quadrature of Dirac operators in Eq.~(\ref{eq:dirac}) and averaging over all states ``$k$'', we obtain the generalized chemical potentials or, for short Klein potentials for electrons $E_e$, neutrons $E_n$ and protons $E_p$
\begin{eqnarray}
\label{Eq:gec1} E_e&=&\sqrt{g_{00}}\mu_e-eV={\rm e}^{\nu/2}\mu_e -eV ,\\
\label{Eq:gec2}E_p &=& \sqrt{g_{00}}\mu_p+g_{\omega}\omega+g_{\rho}\rho+eV ={\rm e}^{\nu/2}\mu_p+g_{\omega}\omega+g_{\rho}\rho+eV, \\
\label{Eq:gec12}E_n&=&\sqrt{g_{00}}\mu_n+g_{\omega}\omega-g_{\rho}\rho ={\rm e}^{\nu/2}\mu_n +g_{\omega}\omega-g_{\rho}\rho,
\end{eqnarray}
where we have used Eqs.~(\ref{Eq:1D})--(\ref{Eq:4D}) and Eqs.~(\ref{eq:dis_k}), (\ref{eq:dis}), (\ref{eq:EOS1})--(\ref{eq:EOS2b}). In the zero-temperature case, they are generalized Fermi energies for electrons $E_e=E_e^F$, neutrons $E_n=E_n^F$ and protons $E_p=E_p^F$.

Using the equations of motion for the fields $\rho$, $\omega$ and $\sigma$, and using the generalized Gibbs-Duhem relation (\ref{Eq:GGDR2}), the energy-momentum conservation equation (\ref{eq:TOV}) can be rewritten as
\begin{eqnarray}\label{eq:TOV1}
{\rm e}^{\nu/2}\sum_{i=n,p,e} n_i \left( d\mu_i - \frac{dT}{T} \mu_i\right) &+& ({\cal E}+{\cal P}){\rm e}^{\nu/2} \left(\frac{dT}{T} + \frac{1}{2} d\nu \right) \nonumber \\
&+& g_{\omega}n_b d\omega +g_{\rho} n_3 d\rho + e n_{ch} dV=0.
\end{eqnarray}
The isothermal Tolman condition \cite{tolman30} (see also \cite{klein49}) demands the constancy of the gravitationally red-shifted temperature
\begin{equation}\label{eq:Tcons}
\frac{dT}{T} + \frac{1}{2} d\nu = 0, \qquad {\rm or}\qquad {\rm e}^{\nu/2} T = {\rm constant}.
\end{equation}
Such a condition can be used into Eq.~(\ref{eq:TOV1}) to obtain
\begin{equation}\label{Eq:TOVPINTRO}
\sum_{i=n,p,e} n_i d({\rm e}^{\nu/2} \mu_i)+ g_{\omega}n_b d\omega +g_{\rho} n_3 d\rho + e n_{ch} dV=0.
\end{equation}
Moreover, using the expressions (\ref{Eq:gec1})--(\ref{Eq:gec2}) of the generalized chemical potentials, Eq.~(\ref{Eq:TOVPINTRO}) can be rewritten as
\begin{equation}\label{Eq:2pas}
\sum_{i=n,p,e} n_i dE_i=0,
\end{equation}
which leads for independent and non-zero particle number densities $n_i \neq 0$ to the constancy of the Klein potentials (\ref{Eq:gec1})--(\ref{Eq:gec12}) for each particle-species, i.e.
\begin{eqnarray}
E_e &=& {\rm e}^{\nu/2}\mu_e -eV={\rm constant},\label{eq:ef1}\\
E_{p} &=& {\rm e}^{\nu/2}\mu_{p} + \mathcal{V}_{p} = {\rm constant},\label{eq:ef2} \\
E_{n} &=& {\rm e}^{\nu/2}\mu_{n} + \mathcal{V}_{n} = {\rm constant},\label{eq:ef3}
\end{eqnarray}
where 
\begin{eqnarray}\label{eq:effpot}
\mathcal{V}_{p} &=& g_{\omega}\omega + g_{\rho} \rho + eV,\\
\mathcal{V}_{n} &=& g_{\omega}\omega - g_{\rho} \rho.
\end{eqnarray}

In the case of nuclear matter in $\beta$-equilibrium (assuming not trapped neutrinos), the values of the constant Klein potentials (\ref{eq:ef1})--(\ref{eq:ef3}) are linked by the condition
\begin{equation}\label{eq:beta}
E_n = E_p + E_e,
\end{equation}
which can be rewritten explicitly in terms of the chemical potentials as
\begin{equation}\label{Eq:3mu}
\mu_n=\mu_p+\mu_e + 2 g_{\rho}\rho {\rm e}^{-\nu/2}.
\end{equation}

%
\section{Concluding Remarks}\label{sec:con}%

We have presented the self-consistent equations of equilibrium at finite temperatures for a system of neutrons, protons
and electrons in $\beta$-equilibrium within the framework of general relativity including quantum statistics, electro-weak, and strong interactions. In the mean-field approximation, we obtained the generalized particle chemical potentials from the Dirac equations for nucleons and electrons.

From the Einstein-Maxwell equations, the thermodynamic laws and energy-momentum conservation, we obtain the constancy of the Klein potential of each particle-specie and of the gravitationally red-shifted temperature throughout the configuration, i.e. the first Klein integrals and the Tolman isothermal condition respectively. In the non-interacting degenerate case, following a minimization energy procedure, it was demonstrated that the thermodynamic equilibrium condition of constancy of the generalized particle Fermi energy of all particle species holds (see E.~Olson and M.~Bailyn \cite{olson75}). Such a procedure can be straightforwardly applied to the present case, being the final result given by the equilibrium conditions (\ref{eq:ef1}) and (\ref{eq:ef2}).

The precise values of such constants are linked, in the case of nuclear matter in $\beta$-equilibrium, by Eq.~(\ref{eq:beta}), and their full determination needs the inclusion of additional constraints to the system, e.g.~global charge neutrality (see e.g.~\cite{PLB2011}).

The correct implementation of such generalized Thomas-Fermi equilibrium conditions needs the self-consistent solution of the global problem of equilibrium of the configuration following from the solution of the Einstein-Maxwell equations (\ref{eq:EM12}), (\ref{eq:EM22}), (\ref{eq:EM32})--(\ref{eq:lambda}), the general relativistic thermodynamic equilibrium conditions (\ref{eq:Tcons}), (\ref{eq:ef1}) and (\ref{eq:ef2}), together with the constraints, e.g. $\beta$-equilibrium and global charge neutrality. 

Thus, the full system of Einstein-Maxwell-Thomas-Fermi equations can be rewritten in the form
\begin{eqnarray}
&&{\rm e}^{-\lambda(r)}\left(\frac{1}{r^2}-\frac{1}{r}\frac{d\lambda}{dr}\right)-\frac{1}{r^2}= -8 \pi G T_0^0, \label{eq:EMTF1}\\
&&{\rm e}^{-\lambda(r)}\left(\frac{1}{r^2}+\frac{1}{r}\frac{d\nu}{dr}\right)-\frac{1}{r^2}= -8 \pi G T_1^1, \label{eq:EMTF2}\\
&&V'' + \frac{2}{r}V' \left[ 1 - \frac{r (\nu'+\lambda')}{4}\right] = - 4 \pi e \, {\rm e}^{\nu/2} {\rm e}^{\lambda} (n_p -n_e) ,\label{eq:GRTF} \\
&&\frac{d^2\sigma}{dr^2}+\frac{d\sigma}{dr}\left[\frac{2}{r}-\frac{1}{2}\left(\frac{d\nu}{dr}+\frac{d\lambda}{dr}\right)\right]={\rm e}^{\lambda}\left[\partial_{\sigma}U(\sigma)+g_s n_s\right], \label{eq:EMTF3}\\
&&\frac{d^2\omega}{dr^2}+\frac{d\omega}{dr}\left[\frac{2}{r}-\frac{1}{2}\left(\frac{d\nu}{dr}+\frac{d\lambda}{dr}\right)\right]=-{\rm e}^{\lambda}\left[g_{\omega}J_{\omega}^0-m_{\omega}^2\omega\right], \label{eq:EMTF4}\\
&&\frac{d^2\rho}{dr^2}+\frac{d\rho}{dr}\left[\frac{2}{r}-\frac{1}{2}\left(\frac{d\nu}{dr}+\frac{d\lambda}{dr}\right)\right]=-{\rm e}^{\lambda}\left[g_{\rho} J_{\rho}^0-m_{\rho}^2\rho\right], \label{eq:EMTF5} \\
&&E_e={\rm e}^{\nu/2}\mu_e -eV={\rm constant},\label{eq:ef12}\\
&&E_p ={\rm e}^{\nu/2}\mu_{p} + \mathcal{V}_{p} = {\rm constant}, \\
&&E_n ={\rm e}^{\nu/2}\mu_{n} + \mathcal{V}_{n} = {\rm constant},\\
&&{\rm e}^{\nu/2} T = {\rm constant},
\end{eqnarray}
where the constants $E_n$, $E_p$ and $E_e$ are linked by Eq.~(\ref{eq:beta}) and $\mathcal{V}_{p,n}$ is given by Eq.~(\ref{eq:effpot}). In particular, in the degenerate case $T=0$, Eq.~(\ref{eq:GRTF}) becomes 
\begin{align}\label{eq:GRTF2}
\hat{V}'' + \frac{2}{r}\hat{V}' \left[ 1 - \frac{r (\nu'+\lambda')}{4}\right] &= - 4 \pi \alpha \, {\rm e}^{\nu/2} {\rm e}^{\lambda} \Bigg\{n_p \nonumber \\
&- \frac{{\rm e}^{-3 \nu/2}}{3 \pi^2}[\hat{V}^2 + 2 m_e \hat{V} - m^2_e ({\rm e}^{\nu}-1)]^{3/2}\Bigg\},
\end{align}
where $\hat{V} \equiv e V + E_e$ and we have used Eq.~(\ref{eq:ef12}) into Eq.~(\ref{eq:GRTF}). This equation is the general relativistic extension of the relativistic Thomas-Fermi equation recently introduced in \cite{PRC2011} for the study of compressed atoms. In addition, Eq.~(\ref{eq:GRTF2}) has been recently used to obtain the globally neutral configurations in the simpler case of degenerate neutrons, protons and electrons in $\beta$-equilibrium (see \cite{PLB2011} for details).

This paper was completed in honor to Oskar Klein in the occasion of the ``weekly Oskar Klein Centre seminar series'' given by one of us (R.~R) at Stockholm University. We would like to thank the referee for the constructive and very interesting suggestions. J.A.R. thanks D.~Pugliese and S.~Cacciatori for helpful discussions on the subject.


\end{document}